\documentclass[%
 reprint,
 amsmath,amssymb,
 aps,
 superscriptaddress,
]{revtex4-2}

\usepackage{graphicx}
\usepackage{dcolumn}
\usepackage{bm}
\usepackage{physics}
\usepackage[colorlinks=true,linkcolor=black,anchorcolor=black,citecolor=black,filecolor=black,menucolor=black,runcolor=black,urlcolor=black]{hyperref}

\begin{document}

\preprint{APS/123-QED}

\title{Resonant and off-resonant magnetoacoustic waves in epitaxial Fe$_3$Si/GaAs hybrid structures}

\author{Marc Rovirola}
\affiliation{Dept. of Condensed Matter Physics, University of Barcelona, 08028 Barcelona, Spain}
\affiliation{Institut de Nanociència i Nanotecnologia (IN2UB), University of Barcelona, 08028 Barcelona, Spain}
\author{M. Waqas Khaliq}
\affiliation{Dept. of Condensed Matter Physics, University of Barcelona, 08028 Barcelona, Spain}
\affiliation{ALBA Synchrotron Light Source, 08290 Cerdanyola del Vallès, Spain}
\author{Blai Casals}
\affiliation{Dept. of Condensed Matter Physics, University of Barcelona, 08028 Barcelona, Spain}
\affiliation{Institut de Nanociència i Nanotecnologia (IN2UB), University of Barcelona, 08028 Barcelona, Spain}
\author{Michael Foerster}
\affiliation{ALBA Synchrotron Light Source, 08290 Cerdanyola del Vallès, Spain}
\author{Miguel Angel Niño}
\affiliation{ALBA Synchrotron Light Source, 08290 Cerdanyola del Vallès, Spain}
\author{Jens Herfort}
\affiliation{Paul-Drude-Institut für Festkörperelektronik, Leibniz-Institut im Forschungsverbund Berlin e.V., Hausvogteiplatz 5-7, 10117 Berlin, Germany}
\author{Joan Manel Hernàndez}
\affiliation{Dept. of Condensed Matter Physics, University of Barcelona, 08028 Barcelona, Spain}
\affiliation{Institut de Nanociència i Nanotecnologia (IN2UB), University of Barcelona, 08028 Barcelona, Spain}
\author{Ferran Macià}
\email{ferran.macia@ub.edu}
\affiliation{Dept. of Condensed Matter Physics, University of Barcelona, 08028 Barcelona, Spain}
\affiliation{Institut de Nanociència i Nanotecnologia (IN2UB), University of Barcelona, 08028 Barcelona, Spain}
\author{Alberto Hernández-Mínguez}
\affiliation{Paul-Drude-Institut für Festkörperelektronik, Leibniz-Institut im Forschungsverbund Berlin e.V., Hausvogteiplatz 5-7, 10117 Berlin, Germany}

\date{\today}

\begin{abstract}

Surface acoustic waves (SAWs) provide an efficient dynamical coupling between strain and magnetization in micro/nano-metric devices. Using a hybrid device composed of a piezoelectric, GaAs, and a ferromagnetic Heusler alloy thin film, Fe$_3$Si, we are able to quantify the amplitude of magnetoacoustic waves generated with SAWs via magnetic imaging in an X-ray photoelectron microscope. The cubic anisotropy of the sample together with a low damping coefficient allows for the observation of resonant and non-resonant magnetoelastic coupling. Additionally, via micromagnetic simulation, we verify the experimental behavior and quantify the magnetoelastic shear strain component in Fe$_3$Si that appears to be very large ($b_2=14\times 10^6$ J/m$^3$), much larger than the one found in Nickel.

\end{abstract}

\maketitle


\section{\label{sec:intro} Introduction}

The use of charge carriers for information transport has proven to be successful, but current technology is reaching its limitations and the growing need for smaller, faster, and more efficient information processing units calls for new data carriers. One of the solutions compatible with existing technology and involving low-energy dissipation are spin waves--- collective excitation of magnetic order. However, generation, detection, and manipulation of spin waves are still the focus of many studies due to the multiple interactions that these magnetization modes have with other degrees of freedom such as phonons or photons \cite{Yi_magnons_JAP2020,Inman_magnons_PRapp2022}.

The coupling between strain and magnetization---the magnetoelastic (ME) effect---can be used to generate magnetization dynamics at the nanoscale with low-power dissipation through surface acoustic waves (SAW) \cite{NatCom2017_Foerster,thevenard2018_pra,MRS2018, MagneticResonanceImaging, casalsGenerationImagingMagnetoacoustic2020, Thevenard_imaging_PRB2020,Labanowskieaat6574}. The interaction between SAW and magnetization dynamics has long been studied \cite{Hernandez2006, Davis2010, Weiler2011, Weiler2012, Ralph2015, Labanowski2016,Thevenard2016, Kuszewski_2018, Labanowskieaat6574, Foerster_2019, reviewSAW_APL_2022} and there are clear experiments showing the variation of magnetic states caused by SAW and, reversibly, changes in SAW propagation are observed due to the action of magnetization dynamics. SAW-induced magnetization waves including magnetoacoustic switching might occur at magnetic resonances \cite{Weiler2011, Weiler2012, Ralph2015,Kuszewski_2018,thevenard2018_pra, Labanowski2016} but there is no clear picture of what type of magnetoacoustic waves might be generated, if any, when magnetic and acoustic systems are in a non-resonance condition.  Imaging techniques capable of quantifying both SAW and magnetization dynamics are required to fully explore the interaction of hybrid modes of magnetoacoustic waves generated by SAW.



The coherent transport of spin waves up to micrometers is a requirement for processing units and for that reason finding materials with low damping coefficients is critical. Materials such as nickel have a large ME effect, which makes them ideal for generating strain-induced spin waves, but display a large damping coefficient in the order of $10^{-3}$ \cite{SchmidtYIGDamping2020} causing shorter propagation distances and broad resonance peaks. On the contrary, Yttrium Iron Garnet (YIG) has a damping coefficient in the order of $10^{-4}$ \cite{SchmidtYIGDamping2020}, but high-quality growth is challenging and it is a weak magnetoelastic material. Epitaxial Heusler alloys are promising alternatives due to their properties, including low spin-wave damping, half-metallicity, and compatibility with CMOS technology \cite{HERFORT2005666,PRBionescu2005,APL_herfort_2003,basAcousticallyDrivenFerromagnetic2020,PRB_gusenbauer_2011,Thomas_2009,PRB_Lenz_2005}. Specifically, Fe$_3$Si films grown epitaxially on GaAs have been reported having a damping coefficient as low as $3\times10^{-4}$ \cite{lenzMagneticAnisotropyResonance2006a} and a large ME effect due to its large magnetoelastic constant $b_2$ \cite{wegscheiderMagneticAnisotropyEpitaxial2011}.


In this article, we study the acoustic excitation of spin waves in samples consisting of a piezoelectric GaAs substrate with interdigital transducers (IDTs) for the generation of SAW, and a ferromagnetic Fe$_3$Si thin film on the SAW acoustic path (see Fig. \ref{setup}). The high structural quality of Fe$_3$Si leads to very narrow ferromagnetic resonance (FMR) peaks indicative of low Gilbert damping ($\alpha\simeq0.005$) and, therefore, to a large coherent length of the spin wave. To quantify the spin waves, direct imaging through X-ray microscopy of both SAW and magnetization waves is taken at different magnetic fields, and we find that magnetoacoustic waves are generated in a wide range of fields, showing that resonance is not an essential condition to generate magnetoacoustic waves. 

The article is organized as follows. Section \ref{section:expdet} describes the sample fabrication process and the experimental procedures. Section \ref{section:results} is divided into two parts: first, the experimental results, beginning with the characterization of the sample via FMR and SAW transmission, and then the analysis of direct imaging of both SAW and magnetic waves. Second, micromagnetic simulations are shown to further verify and quantify the magnetoelastic constants of the sample and support the idea that magnetoacoustic waves can also exist at non-resonance conditions. Lastly, Section \ref{section:conclusions} presents the conclusions on the results.

\section{Experimental details}\label{section:expdet}
\begin{figure*}[t]
    \centering
    \includegraphics[width=0.9\textwidth]{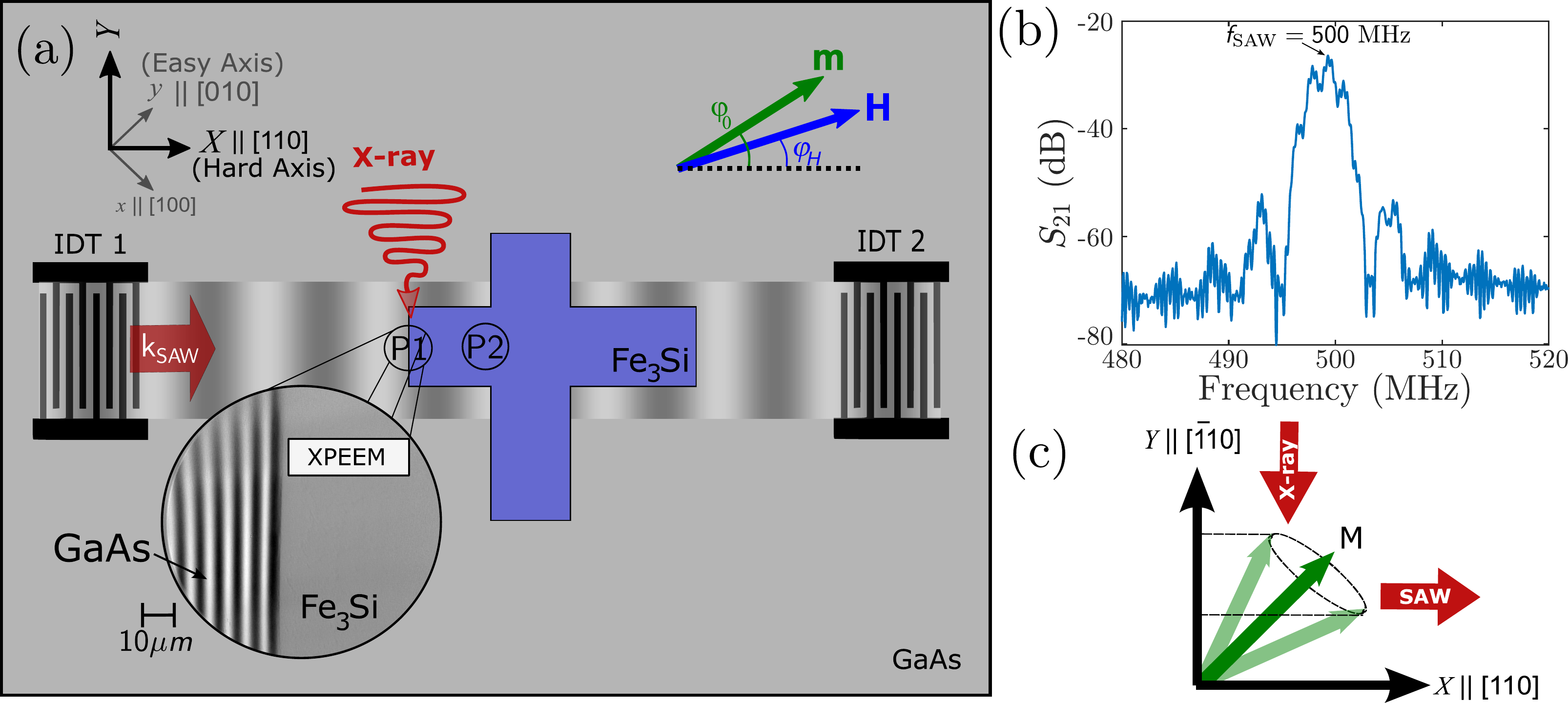}
    \caption{ \textbf{(a)} 	Schematic set-up of the hybrid device, where a cross-shaped Heusler alloy film (Fe$_3$Si) is deposited on top of a GaAs substrate with interdigital transducers (IDTs) on both sides. The top left corner shows the coordinates of the sample, where $x$ and $y$ are the easy axes, while $X$ and $Y$ are the hard axes. IDT 1 is excited with an RF signal at 500 MHz to excite SAW and the X-rays are sent perpendicular to the SAW propagation direction. The top right corner shows the equilibrium magnetization and magnetic field directions with angles $\varphi_0$ and $\varphi_{\mathrm{H}}$ respectively. \textbf{(b)} Transmission coefficient between the two IDTs, $S_{21}$,  as a function of the RF frequency, applied to IDT 1. The arrow points to the highest transmission and fundamental frequency of IDT 1 at 500 MHz. \textbf{(c)} Schematic representation of the magnetization precession in the Fe$_3$Si film with the SAW propagating in the $X$ direction and the X-rays in the $Y$ direction. The dashed lines that stretch in the $Y$ direction show the magnetic signal that is captured by XMCD. }
    \label{setup}
\end{figure*}

The studied sample is a 74-nm-thick Fe$_3$Si film grown by molecular beam epitaxy on a piezoelectric, GaAs (001)\cite{HERFORT2005666}. The fabrication of the magnetoacoustic device begins by selective patterning of the Fe$_3$Si film into cross-shaped structures by electron beam lithography and plasma etching. Pairs of IDTs are deposited on opposite sides of patterned Fe$_3$Si film by electron beam lithography, metal evaporation, and liftoff (see Fig.\ \ref{setup}), thus allowing for both the generation of SAWs that propagate along the Heusler film and the measurement of transmitted SAW power ($S_{21}$) along the acoustic path. Figure \ref{setup}(b) displays $S_{21}$ spectrum measured with a vector network analyzer. IDTs are designed to excite Rayleigh SAW \cite{SAW1,SAW2} with a frequency of $f_{\text{SAW}}=500$ MHz and the corresponding wavelength $\lambda_{\text{SAW}}=5.73$ $\mu$m propagating along $X\! \parallel \! [110]$ direction of the GaAs substrate. They are characterized by longitudinal in-plane and out-of-plane strains $\epsilon_{XX}$ and $\epsilon_{ZZ}$, respectively, as well as shear strain $\epsilon_{XZ}$. Due to the large structural quality of the Fe$_3$Si film, the magnetization displays cubic in-plane anisotropy, with easy and hard axes being the family of planes $\expval{100}$ and $\expval{110}$, respectively. Therefore, the SAW propagates along one of the hard axes of the Fe$_3$Si film, and magnetization in the absence of a magnetic field lies along any of the four easy axes forming 45 degrees with respect to the SAW path.

To image the sample, we used an X-ray photoemission electron microscopy (XPEEM) set-up from the CIRCE beamline at ALBA Synchrotron with a time and spatial resolution below 80 ps and 100 nm, respectively. The SAW frequency is synchronized to the repetition rate of the X-ray pulses, and the SAW and magnetization images are taken simultaneously \cite{Foerster2016,Foerster_JSR,NatCom2017_Foerster,casalsGenerationImagingMagnetoacoustic2020}. The sample was mounted on a holder containing an electromagnet that can generate an in-plane magnetic field up to $20\text{ mT}$ in the SAW propagation direction. Sending an RF signal of the appropriate frequency to IDT\,1 generates SAW in the GaAs that may excite spin waves in the Fe$_3$Si with the same wavelength. When the sample and the electron microscope are kept at a large potential difference, electrons leaving the sample due to X-rays are accelerated to the electron microscope, but the amount of electrons captured from each position depends on the amplitude of the piezoelectric field accompanying the SAW \cite{Foerster_JSR}. Therefore, we obtain contrast images sensitive to SAW wavefronts, as seen in the inset of Fig.\ \ref{setup}. No SAW can be detected in Fe$_3$Si due to its metallic nature, which shields the dynamic piezoelectric field. However, the effect of SAW on the magnetization in the Fe$_3$Si can be imaged using X-ray magnetic circular dichroism (XMCD), which consists of subtracting two otherwise identical XPEEM images using X-rays with opposite circular polarization \cite{Stohr2007}.

\section{Results}\label{section:results}
\subsection{Experimental results}\label{section:expresults}
\begin{figure}
    \centering
    \includegraphics[width=1\columnwidth]{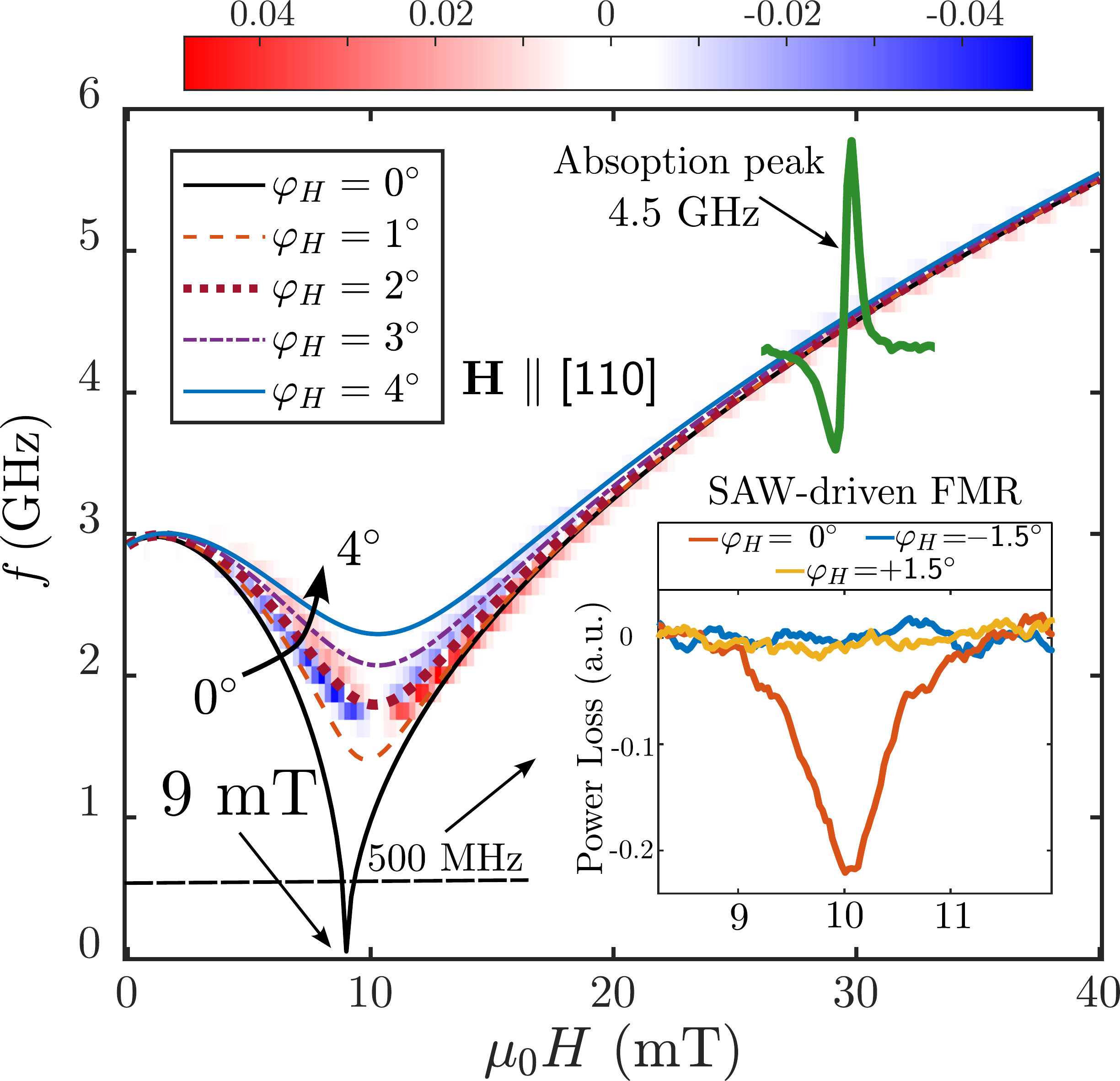}
    \caption{FMR for a magnetic field applied along the [110] crystalline direction (hard axis) with the modeling curve plotted on top as continuous, dashed, and dotted lines for different angles of the magnetic field, $\varphi_H$. The horizontal dashed line corresponds to the SAW frequency and the arrow points to the anisotropy field (9 mT) obtained with the model. The green curve at 4.5 GHz displays the absorption peak in the form of a Lorentzian derivative. The inset shows SAW attenuation at 500 MHz as a function of the magnetic field strength for different field angles, $\varphi_{\mathrm{H}}$ between SAW and H.}
    \label{FMR}
\end{figure}

The sample was initially characterized via FMR with a broadband coplanar waveguide capable of producing microwave magnetic fields with frequencies up to 20 GHz \cite{vegard_prb_2015}. Figure \ref{FMR} shows the resonance fields for the uniform precession mode (spin wave with $\textbf{q}=0$) when an in-plane magnetic field is applied in the [110] direction (hard axis). On top of the data, we plotted solid and dashed lines calculated from a macrospin model at several magnetic field alignments. The model accounts for the fourfold cubic crystalline anisotropy, and the shape anisotropy, see Refs. \cite{PRB_Lenz_2005, Alberto_2020}. We notice here that a curve corresponding to a magnetic field direction that slightly deviates $\varphi_{\mathrm{H}} \simeq 2^\circ$ from [110] axis fits well the experimental data. The curve corresponding to a perfectly aligned magnetic field, $\varphi_{\mathrm{H}} = 0^\circ$ (continuous black curve), clearly shows the value of the anisotropy field, $\mu_0 H_k = 9$ mT, that corresponds to the magnetic field required to align the magnetization along the hard axis (i.e., the magnetic field required to suppress the effective magnetic field). We also measured the linewidth broadening as a function of microwave frequency and found a value for the Gilbert damping coefficient, $\alpha = 0.005$ \cite{suppmat1}, an order of magnitude smaller than Nickel \cite{walowskiIntrinsicNonlocalGilbert2008}. 

We plotted in Fig.\ \ref{FMR} an additional dashed line representing the frequency used in the SAW experiment (500 MHz). The crossing points between the FMR curves and the dashed line indicate the resonance fields at which there is a maximum energy transfer between microwave magnetic fields (from the coplanar waveguide) and the film's magnetization. Small deviations in the applied magnetic field direction result in an FMR curve shift and no intersection points---or no resonance condition \cite{suppmat2}. This effect is studied in detail later on with micromagnetic modeling. 




To further characterize our hybrid structures, we measured SAW attenuation as a function of the applied magnetic field along the [110] direction (see the inset of Fig.\ \ref{FMR}). The measurement of power transmission, $S_{21}$, between two opposite IDTs serves to investigate the ME coupling between SAW and spin waves in our Heusler thin film. Similar to the FMR measurements, there is a sharp absorption peak with a linewidth of approximately $1 \text{ mT}$ and a strong angular dependence \cite{Alberto_2020}. A deviation of $\varphi_{\mathrm{H}} = 1.5^\circ$ of the magnetic field direction with respect to the [110] axis results in no measurable power loss (see Fig. \ref{FMR} inset)
 
Next, we turn to the main experimental results consisting of resolving in space and time the evolution of magnetoacoustic waves in the Heusler film. First, we characterized the strength of a 500 MHz SAW by direct XPEEM images of the GaAs substrate. The observed SAW wave fronts in the XPEEM image (see the left-hand side of the inset in Fig. \ref{setup}) are associated with the energy shift of secondary electrons leaving the sample surface of GaAs due to the piezoelectric field of the SAW. The energy shift can be indeed quantified via local photo-electron kinetic energy scan---photoemission intensity as a function of the electron kinetic energy---\cite{Foerster_JSR}. The difference between both spectra taken at the wave edges (the difference between maxima and minima) corresponds to the peak-to-peak value and was found to be $V_{pp} = 0.1\text{ V} $. By numerically solving the coupled elastic and electromagnetic equations for a SAW propagating along the [110] axis of GaAs \cite{foersterQuantificationPropagatingStanding2019}, we estimate that the measured piezoelectric amplitude of $V_0 \approx V_{pp}/2 \approx 0.05$ V corresponds to strain components with amplitudes $\varepsilon_{\mathrm{XX}} = 7 \cdot 10^{-5}$, $\varepsilon_{\mathrm{ZZ}} = 3 \cdot 10^{-5}$ and $\varepsilon_{\mathrm{XZ}} = 1 \cdot 10^{-7}$ at the surface of the sample. This is almost an order of magnitude smaller than the same quantity measured in LiNbO$_3$ piezoelectric structures under similar SAW excitations.


\begin{figure}
    \centering
    \includegraphics[width=1\columnwidth]{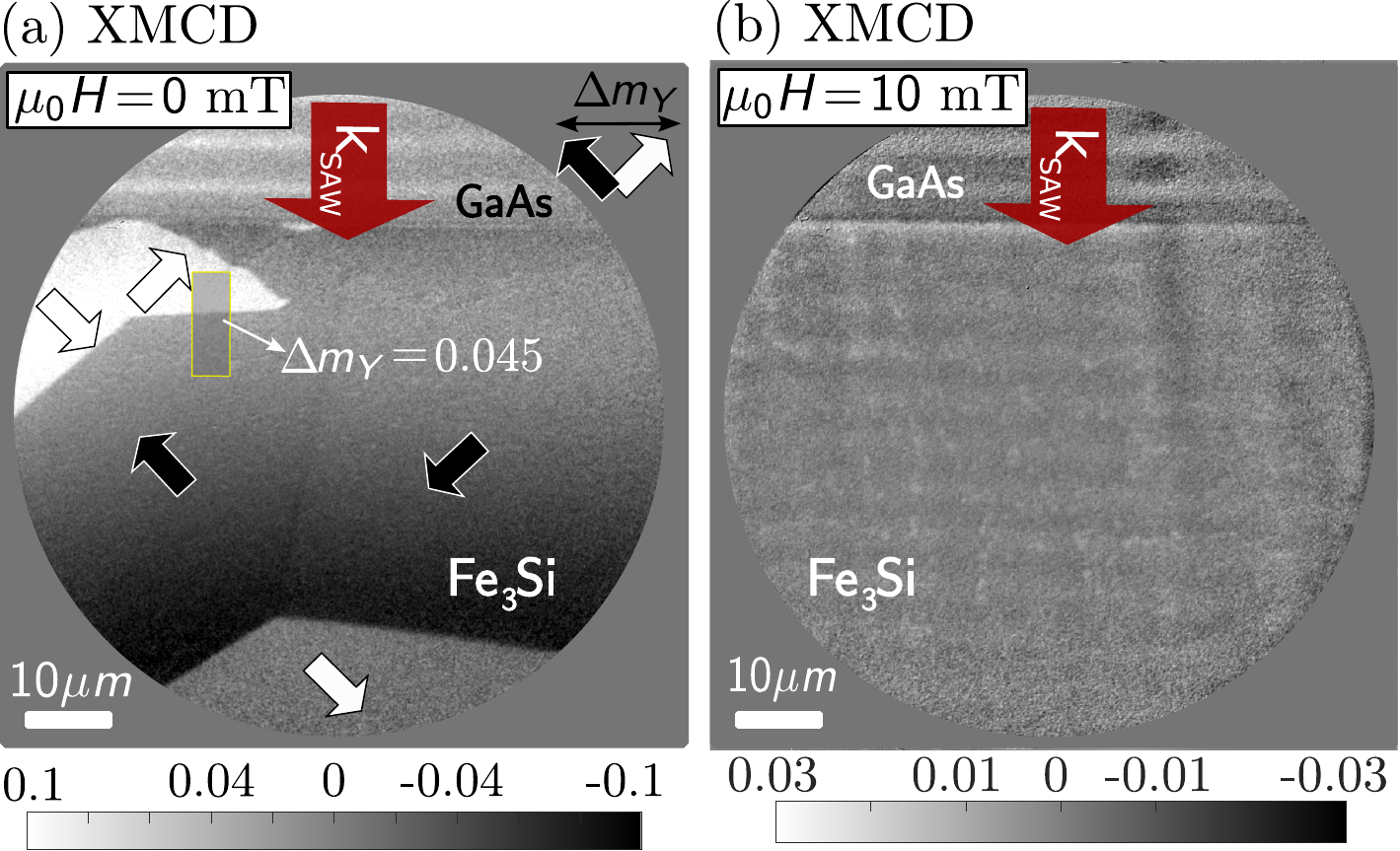}
    \caption{(a) XMCD image of the magnetic domains in Fe$_3$Si and of the SAW in the GaAs. The magnetization direction of each domain is represented by white and black arrows. The difference between opposing magnetic states is represented by $\Delta m_Y$. (b) XMCD image with SAW in the GaAs and spin waves in Fe$_3$Si.}
    \label{Domain_XMCD}
\end{figure}

To obtain XMCD images that are proportional to the Fe$_3$Si magnetization component along the X-ray incidence direction, we selected the X-ray energy for the $L_3$ absorption edge of Fe \cite{Stohr2007}. In our experiment, the direction of incidence of the X-ray is along the hard axes ($Y$ axis) perpendicular to the SAW propagation direction (see Fig.\ \ref{setup} (c)). Figure \ref{Domain_XMCD}(a) shows an XMCD image revealing magnetic domains in all four easy axis directions with a strong contrast that allows us to quantify the overall change in magnetization in the X-ray direction.


SAWs create a dynamic strain in the Fe$_3$Si film that induces a time-varying magnetic anisotropy, which eventually creates magnetoacoustic waves. The amplitude of these magnetoacoustic waves can be seen in the XMCD image, but with a signal of 1-2 orders of magnitude smaller compared to the magnetic domains observed in Fig.\ \ref{Domain_XMCD}(a). Figure \ref{Domain_XMCD}(b) displays an XMCD image under a magnetic field of $\mu_0 H \simeq$ 10 mT, showing magnetoacoustic waves in the Fe$_3$Si film (there are no magnetic domains for applied fields larger than 1 mT). A remaining SAW signal in the GaAs is still visible, both in Fig.\ \ref{Domain_XMCD}(a) and (b), due to tiny temperature drifts during integration time, which make the two XPEEM images taken for XMCD slightly different and thus not fully canceling out the SAW component.

\begin{figure}
    \centering
    \includegraphics[width=1\columnwidth]{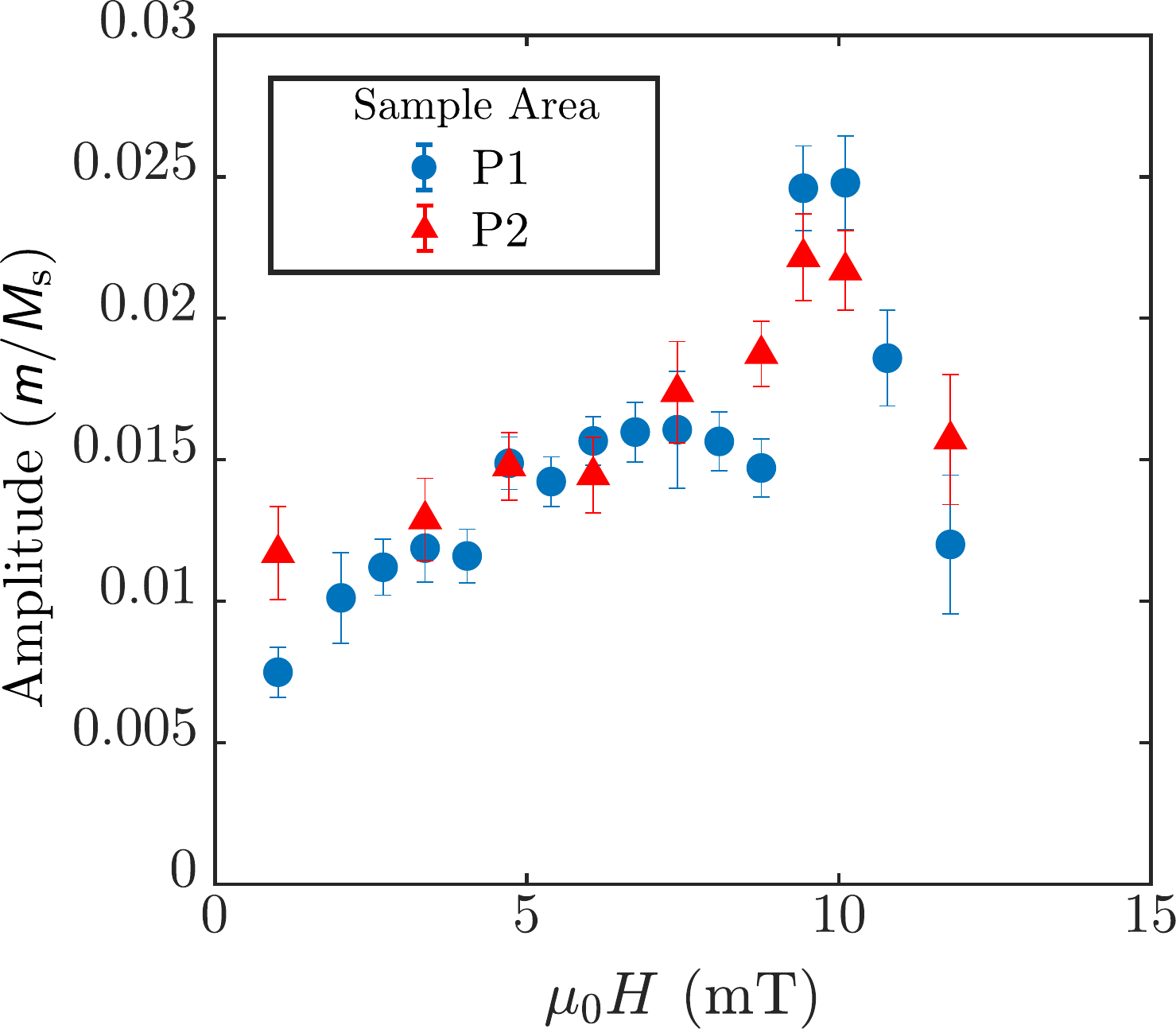}
    \caption{
    Spin wave amplitude obtained from the XMCD images as a function of the in-plane magnetic field at two different positions of the sample as indicated in Fig. \ref{setup} (a).}
    \label{ALBAResults}
\end{figure}

To study the importance of magnetic resonance condition in the generation of magnetoacoustic waves, we measured XMCD images as a function of the applied magnetic field similar to what we did in the FMR and SAW transmission characterization in Fig.\ \ref{FMR}. We took XMCD images at two different areas in the sample, P1 and P2, as shown in Fig.\ \ref{setup}. Location P1 includes GaAs substrate and the Fe$_3$Si film as seen in Fig.\ \ref{Domain_XMCD}(b). By fitting the magnetoacoustic wave profile to a sinusoidal function \cite{suppmat3}, we can retrieve the amplitude and quantify it with respect to the overall $M_s$ value found from the magnetic-domain image \cite{suppmat4}. The normalized amplitude is showcased in Fig.\ \ref{ALBAResults} for the two areas and shows that, contrary to FMR and SAW transmission measurements, there are sizable magnetoacoustic waves on a large magnetic field range with a sharp increase around 10 mT, which is exactly the magnetic resonance field for spin waves with $\textbf{q}=\textbf{k}_{\mathrm{SAW}}$. Thus, magnetoacoustic waves are created with SAW even when there is no measurable power loss in FMR or SAW transmission---and are excited more efficiently at magnetic resonance. Our quantification of the magnetoacoustic waves gives an estimate of 0.5-1.5 degree precession of amplitude for the magnetization vector around the equilibrium.



The results of both sample locations are similar and are plotted in Fig.\ \ref{ALBAResults}. Two parts can be identified: the first resembles a broad inverted parabola with its vertex close to 9 mT, and the second is a peak of 1 mT width close to 10 mT. The first part corresponds to non-resonant magnetoacoustic excitation. In the absence of an applied magnetic field, the effective field associated with the cubic anisotropy keeps the magnetization along the in-plane easy axes of the Fe$_3$Si film (as seen in the XMCD image of Fig. \ref{Domain_XMCD}(a)). As the in-plane magnetic field increases, it generates a torque that partially counteracts the anisotropy, until it reaches the anisotropy field of 9 mT. At this point, the total effective field is zero, so that the magnetization can easily respond to the effective magnetic anisotropy created by the SAW. Increasing further the applied magnetic field creates again an effective magnetic field that locks the magnetization so that oscillations around the equilibrium direction become smaller. The second part of the curve corresponds to resonance or near-resonance effects close to 10 mT where there is an increase in the amplitude of the magnetoacoustic waves with a width of 1 mT, matching the width of the power loss at magnetic resonance in both FMR and SAW transmission (see Fig.\ \ref{FMR}).

\subsection{Micromagnetic simulation}\label{section:simulations}

We performed micromagnetic modeling using numerical simulations with MuMax3 \cite{mumax3} to estimate magnetoelastic constants, $b_1$ and $b_2$, that are associated with the interaction strength between the strain of the SAW and the magnetization of the Fe$_3$Si. We must also check whether the microscopic model predicts the magnetoacoustic excitations at both resonance and non-resonance magnetic fields. The modeling accounts for all magnetic interactions, including dipolar, cubic anisotropies, and exchange interactions together with time and spatially oscillating strain fields that simulate a SAW. We used the strain field amplitudes found in Sec.\ \ref{section:expresults} and the following values for the ferromagnetic Fe$_3$Si film found via FMR: $M_s = 955 \times 10^3$ A/m, $A_{ex} = 1\times10^{-11}$ J/m, $\alpha = 0.005$  and $K_{c1}=-4400$ J/m$^3$, where $M_s$ is the saturation value of magnetization, $A_{ex}$ is the exchange stiffness and $K_{c1}$ is the first-order anisotropy constant. The value of $A_{ex}$ was calculated using the exchange stiffness constant $D=240$ meV $\r{A}^{2}$ \cite{szymanskiSpinDynamicsFe3Si1991} and the relation $A_{ex}=DM_s/2g\mu_{\mathrm{B}}$ where $g$ is the Landé factor and $\mu_{\mathrm{B}}$ is the Bohr magneton. The free energy of the magnetoelastic interaction can be written as:

\begin{align}
   \nonumber F_{\mathrm{me}}= &\,\, b_1\left[\varepsilon_{x x} m_x^2+\varepsilon_{y y} m_y^2+\varepsilon_{z z} m_z^2\right] \\
    &+2 b_2\left[\varepsilon_{x y} m_x m_y+\varepsilon_{x z} m_x m_z+\varepsilon_{y z} m_y m_z\right],
\end{align}

\noindent where $b_1$ and $b_2$ are magnetoelastic constants. The strain and magnetization components are expressed in the coordinate system $(x,y,z)$ with axes parallel to the $\expval{100}$ directions of the GaAs and Fe$_3$Si. The SAW strain components $\varepsilon_{XX}$, $\varepsilon_{ZZ}$ and $\varepsilon_{XZ}$ are related to the strain components in the coordinate system $(x,y,z)$ using the following formulas:

\begin{equation}
\begin{array}{rl}
\varepsilon_{xx}=\varepsilon_{XX} \cos ^2\theta, & \varepsilon_{xy}=\varepsilon_{XX} \cos \theta \sin \theta, \\
\varepsilon_{yy}=\varepsilon_{XX} \sin ^2\theta, & \varepsilon_{xz}=\varepsilon_{XZ} \cos \theta, \\
\varepsilon_{zz}=\varepsilon_{ZZ}, & \varepsilon_{yz}=\varepsilon_{XZ} \sin \theta,
\vspace{5mm}
\end{array}
\end{equation}

\noindent where $\theta$ is the angle between the SAW propagation direction and the coordinate system defined by the lower case letters $(x, y, z)$. The SAW strain is modeled as a propagating wave that depends on the amplitude and propagation direction as:
\begin{equation}
    e_{jk} = \varepsilon_{jk}\exp({i( \textbf{k}\textbf{r}- wt)})
\end{equation}

\begin{figure}
    \centering
    \includegraphics[width=1\columnwidth]{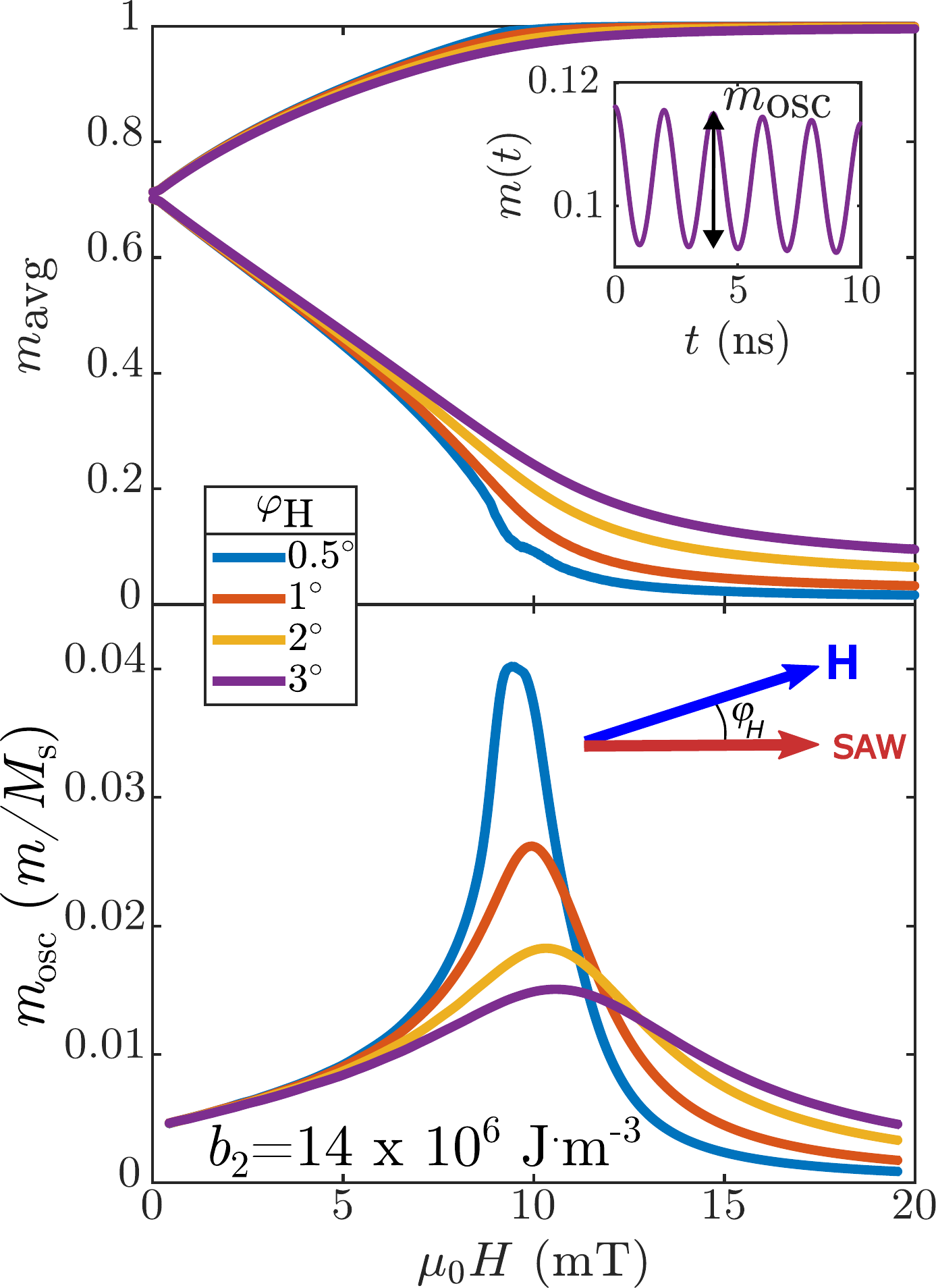}
    \caption{\textbf{(a)} Normalized magnetization components $m_X$ and $m_Y$ as a function of the magnetic field strength for several degrees of misalignment between the magnetic field and the SAW propagation direction. \textbf{(b)} Amplitude of the normalized magnetic component $m_Y$ (spin wave amplitude) as a function of the magnetic field for different angles $\varphi_{\mathrm{H}}$.}
    \label{SimulationResults}
\end{figure}

\noindent where $\varepsilon_{jk}$ is the amplitude, $\textbf{k}$ and $w$ are the wave vector and the angular frequency of the SAW, respectively. We varied the strength of the applied magnetic field as well as the angle $\varphi_{\mathrm{H}}$ from $0.5^\circ$ up to $3^\circ$. We also took the magnetoelastic constants $b_1$ and $b_2$ as parameters.

The results of the simulation are summarized in Fig.\ \ref{SimulationResults}, where the evolution of both the equilibrium magnetization and the oscillation amplitudes are plotted as a function of the applied magnetic field. Figure \ref{SimulationResults}(a) shows the normalized magnetization components $m_X$ and $m_Y$ and shows that, in the absence of an applied magnetic field, the magnetization remains along the easy axis, with each component having an amplitude of $m_{X}=m_{Y}=\cos(45^\circ)$. As the magnetic field increases, the magnetization rotates toward the magnetic field (parallel to the $X$ axis) until it saturates at the anisotropy field (9 mT). Figure \ref{SimulationResults}(b) shows the amplitude of the oscillatory component $m_{\text{osc}}$---equivalent to the component extracted from the waves in the XMCD images---, at different angles of misalignment, $\varphi_{\mathrm{H}}$, between SAW and applied magnetic field. The amplitude of the magnetization oscillation has the same behavior as the one found from the XMCD experiment: a broad peak centered at the anisotropy field $\mu_0 H_k=9$ mT and a narrow 1 mT peak centered at magnetic resonance close to $\mu_0 H=10$ mT. We notice that as the angle of misalignment, $\varphi_{\mathrm{H}}$, increases, the resonance peak disappears, and only the broad peak is left. The simulations for a misalignment of $1-2^\circ$ match well with our data from the XMCD-XPEEM experiment.

Finally, we investigated the dependence of the simulated curves on the values of the magnetoelastic constants $b_1$ and $b_2$. We found that $b_1$ has little effect on the amplitude of the magnetic oscillations unless $b_2$ is small. Therefore, we fixed $b_1=5\times10^6$ J/m$^3$ and varied $b_2$ in the range between 0 and $15\times10^6$ J/m$^3$. The best agreement between simulations and experimental results are the ones shown in Fig. \ref{SimulationResults}(b) and are obtained for $b_2=14\times10^6$ J/m$^3$.

\section{Conclusions}\label{section:conclusions}
We have directly imaged and quantified magnetoacoustic waves in Fe$_3$Si generated with SAW. We observed that magnetoacoustic waves are generated within a wide range of applied magnetic fields with a clear boost at magnetic resonance. The boost at the resonance field matches the FMR and SAW-FMR absorption peaks indicating energy transfer between the SAW and the magnetic system. However, we clearly observed magnetoacoustic waves at non-resonance fields---with little energy absorption. Thus, we conclude that resonance and energy absorption techniques do not provide a full picture of existence of magnetoacoustic waves in magnetic materials. XPEEM images allowed us to quantify the SAW strain components, and the amplitude of the SAW-induced spin waves was obtained from the corresponding XMCD images. Finally, through micromagnetic simulations, we have determined the strength of the shear magnetoelastic constant  ($b_2=14\times 10^6$ J/m$^3$) which is much larger than the one found in Nickel ($b_{2(Ni)}=4.3\times 10^6$ J/m$^3$ \cite{liSpinWaveGeneration2017b}).

\begin{acknowledgments}
The authors would like to thank H.P. Schönherr and S. Meister for their technical support in the preparation of the samples. MR, BC, JMH, MWQ and FM acknowledge funding from MCIN/AEI/10.13039/501100011033 through grant number: PID2020-113024GB-100. MF and MAN acknowledge funding from MCIN through grant numbers: RTI2018-095303 and PID2021-122980OB-C5. MWK acknowledges also Marie Skłodowska-Curie grant agreement No. 754397 (DOC-FAM) from EU Horizon 2020.
\end{acknowledgments}






\bibliography{apssamp}

\end{document}


\title{Supplementary Material}

\begin{center}
    \textbf{\large Supplementary Material}
\end{center}

\section{Supplementary Note 1: Ferromagnetic Resonance Measurements}

\begin{figure}
\includegraphics[width=1\columnwidth]{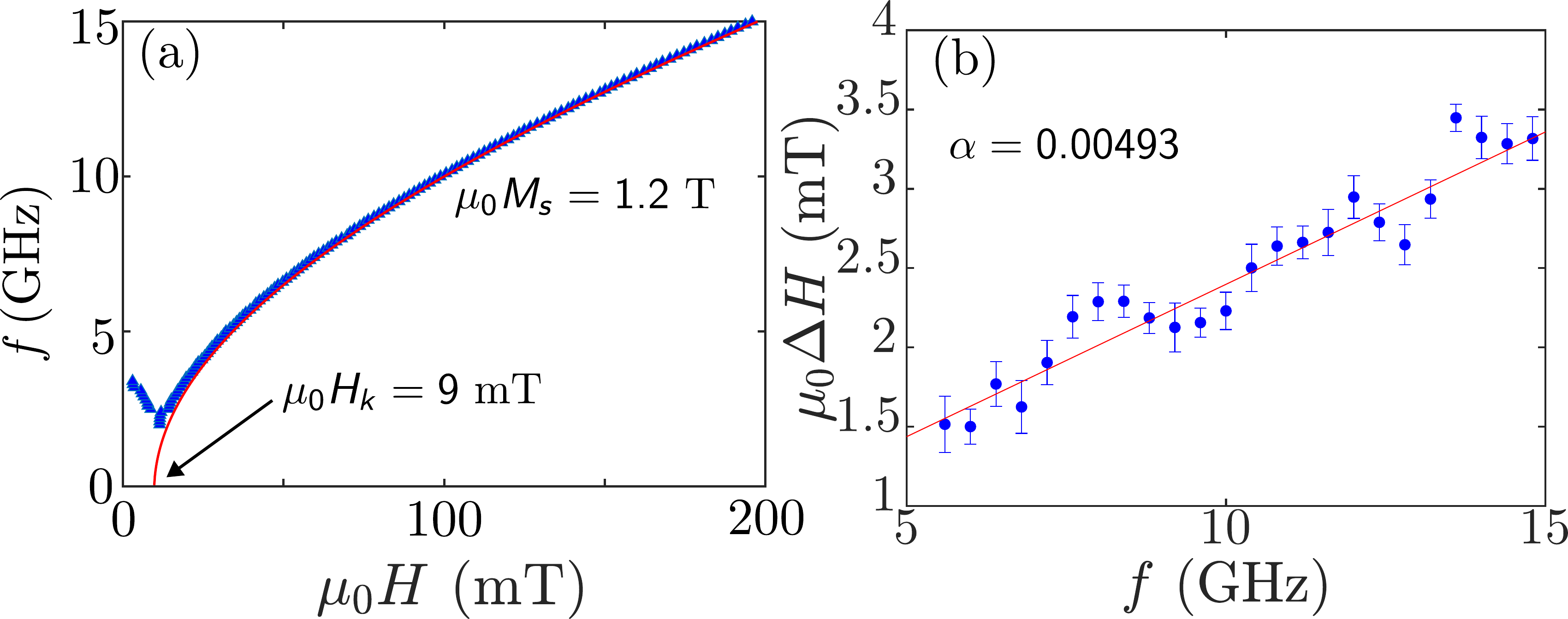}
\caption{(a) Resonant frequency for each magnetic field. The red line is a fit to the data using Kittel's formula (Eq. \ref{kittel}). The arrow points to the anisotropy field of $\mu_0 H_k = 9$ mT. (b) Line width as a function of the resonance frequency. The red line represents the fit from Eq. \ref{damping}, from which we obtain a Gilbert damping parameter of 0.00493.}
\label{FMRSuppMat}
\end{figure}

The Ferromagnetic Resonance (FMR) spectroscopy measurements consisted of placing the Fe$_3$Si sample on a coplanar waveguide and between the poles of an electromagnet. By connecting the coplanar waveguide to a Vector Network Analyzer (VNA) we applied ac electrical currents, ranging from 100 MHz to 15 GHz, that generate an oscillating magnetic field perpendicularly to the magnetic field from the electromagnet. The magnetic moment of the Fe$_3$Si precess in response to the oscillating fields showing energy absorption at resonance fields, which are set by the internal magnetic field, which includes anisotropy, dipolar fields, and an externally applied field. When the frequency of the oscillating field matches the uniform precession mode ($\textbf{q} = 0$) frequency of the spins, the system is at resonance and the energy of the ac signal is absorbed by the Fe$_3$Si, resulting in an absorption peak at the transmission curve ($S_{21}$). These peaks are fitted to a Lorentzian derivative function to find the peak center (resonant field) and the peak width, which is directly related to the Gilbert damping. Figure \ref{FMRSuppMat} (a) shows the resonance frequency as a function of the magnetic field applied along the [110] direction and Figure \ref{FMRSuppMat} (b) shows the width of the peak as a function of the frequency. 

Fitting the resonance values of Fig.\ \ref{FMRSuppMat} (a) to Kittel's formula:

\begin{equation}\label{kittel}
    f = \gamma \mu_0 \sqrt{(H_{0} - H_{k})(H_{0} - H_{k} + M_{s})},
\end{equation}

\noindent we find the anisotropy field, $\mu_0 H_{k} = 9$ mT, and the magnetization saturation, $\mu_0 M_s = 1.2$ T, which corresponds to $M_{\mathrm{s}} = 955\times 10^{3}$ A/m. From $\mu_0 H_{k}$, we can extract the first-order cubic anisotropy $K_{c1} = \mu_0 H_{k} M_s/2 \simeq 4400$ J/m^3.

The Gilbert damping parameter, $\alpha$, can be found by fitting the peak width as a function of the frequency according to the formula:
\begin{equation}\label{damping} 
    \mu_0 \Delta H = \frac{2}{\gamma} \alpha f + \mu_0\Delta H_0,
\end{equation}
\noindent where $\mu_0 \Delta H$ is the full width at half maximum (FWHM). Figure \ref{FMRSuppMat} (b) shows the result of the fit from which we obtained a value of $\alpha \simeq 0.005$. We notice that the inhomogeneous damping, $\mu_0\Delta H_0$, which is the value of the fitted peak width at zero frequency is about 0.5 mT, indicating a good film quality. The values found in this experiment are then used in the MuMax3 simulations.

\section{Supplementary Note 2: Spin wave resonance curve}
Exciting the Fe$_3$Si with SAW generates spin waves with a finite wavenumber $k=2\pi / \lambda_{\mathrm{SAW}}$, where $\lambda_{\mathrm{SAW}}$ is the SAW wavelength. The resonance frequencies of these spin waves differ from the ones excited by the coplanar waveguide, which correspond to the uniform precession mode ($k=0$). Figure 2 shows the resonance frequency of both spin wave modes as a function of the magnetic field strength. For magnetic fields above the anisotropy field (9 mT), the frequency difference is negligible, especially for small magnetic field angles around 500 MHz.

\begin{figure}[!h]
    \centering
    \includegraphics[width=0.8\columnwidth]{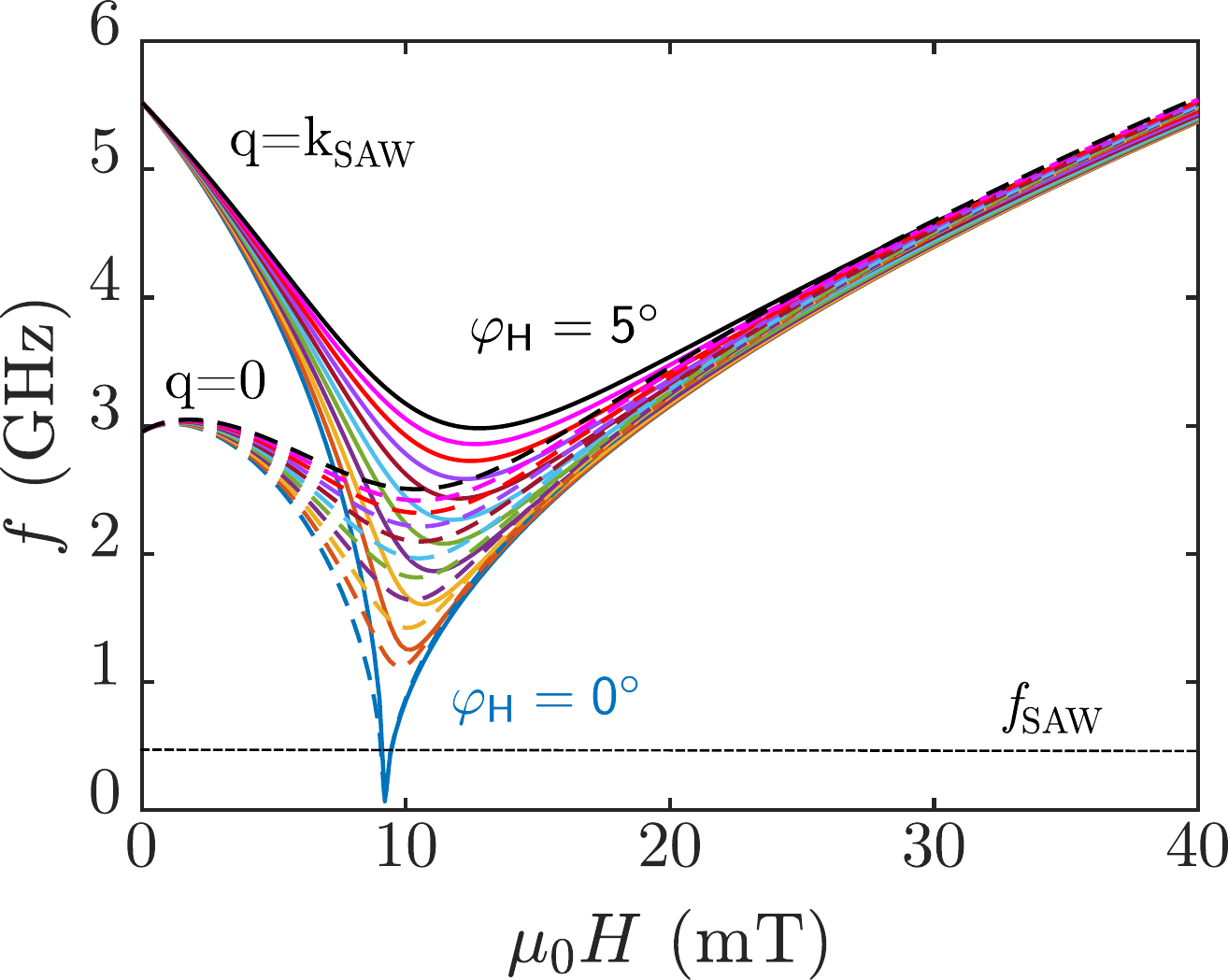}
    \caption{Resonance curves for spin waves with $q=k_{\mathrm{SAW}}$ (solid lines) and $q=0$ (dashed lines), calculated for several magnetic field alignments from 0$^\circ$ to 5$^\circ$ with 0.5$^\circ$ intervals. The straight dashed line represents the SAW frequency of $f_{\mathrm{SAW}}$ = 500 MHz at which the difference between spin wave modes is small.} 
    \label{SpinWaveFMR}
\end{figure}

\vspace{100pt}
\section{Supplementary Note 3: XMCD Data Analysis}
\begin{figure}
    \centering
    \includegraphics[width=1\columnwidth]{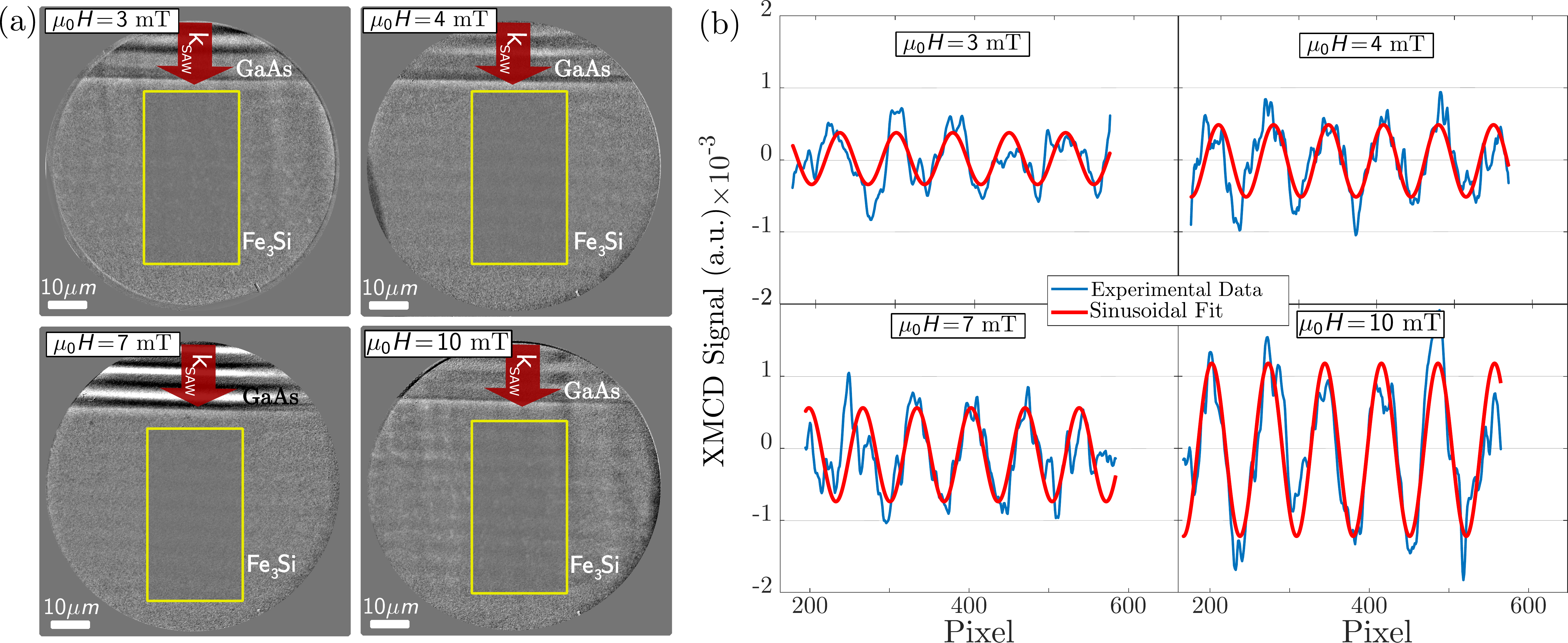}
    \caption{XMCD images with GaAs on the top part and Fe$_3$Si on the bottom part for different applied magnetic fields: 3, 4, 7 and 10 mT. (b) spatial dependence of the XMCD signal along the regions marked with the yellow box in panel (a). The red curves are fittings to the data using Eq. \ref{sinFit}.}
    \label{XMCDSuppMat}
\end{figure}

To analyze the XMCD images, we take an area of the Fe$_3$Si region, as shown in Fig.\ \ref{XMCDSuppMat}(a) and average each row to get the  spin wave patterns displayed in Fig. \ref{XMCDSuppMat}(b). From these traces, we can obtain the amplitude of the spin wave in units of the XMCD signal. The sinusoidal curve fit is as follows:
\begin{equation}\label{sinFit}
    A \sin(kx+\phi_0),
\end{equation}

\noindent where $A$ is the amplitude, $k=2\pi/\lambda$ is the wave number, $x$ is the pixel position and $\phi_0$ is the phase with respect to the edge of the sample or a common defect appearing in all XPEEM images. Figure \ref{XMCDSuppMat}(b) shows the sinusoidal fits for several magnetic fields. The amplitude of the spin wave increases as the magnetic field approaches the anisotropy field, and at 10 mT, there is a jump in amplitude as the system enters resonance (or near-resonance).

\section{Supplementary Note 4: XMCD signal to normalized magnetization}
\begin{figure}[h]
    \centering
    \includegraphics[width=1\columnwidth]{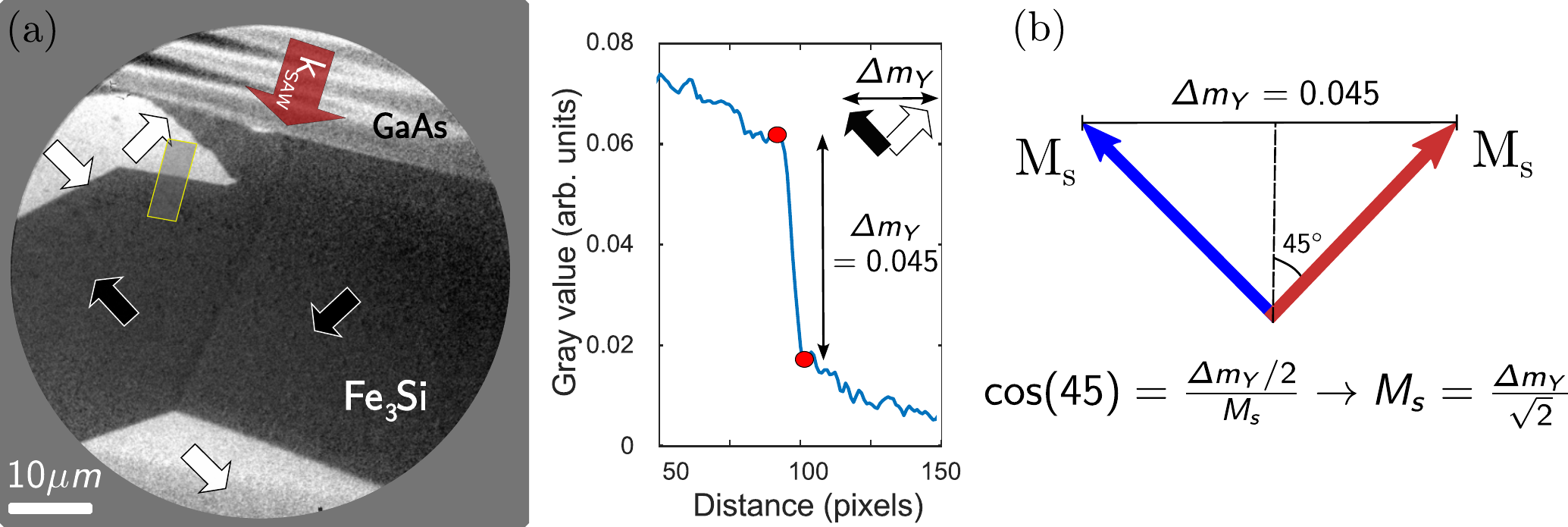}
    \caption{Schematic of magnetization state of domains with a difference of $\Delta m_Y$ to find the normalization value.}
    \label{normSuppMat}
\end{figure}

To convert from an XMCD signal to a normalized magnetization, we use a magnetic domain image (taken at $\mu_0 H = 0$ mT) and calculate the signal difference between two magnetic domains pointing along opposite directions with respect to the $Y$ axis (marked with a rectangle in Fig.\ \ref{normSuppMat}(a)). The magnetization, in absence of a magnetic field, lies in the easy axes; therefore, the two states can be 90 or 180 degrees apart. Figure \ref{normSuppMat}(a) shows the difference between two magnetization states that are 90 degrees apart with a value of $\Delta m_Y = 0.045$. Figure \ref{normSuppMat}(b) shows the steps to obtain $M_{\mathrm{s}}$ in XMCD units.

\title{Supplementary Material}

\begin{center}
    \textbf{\large Supplementary Material 2}
\end{center}
\section{Micromagnetic simulation code (MuMax3)}
\begin{Verbatim}[baselinestretch=0.95]
// The following code shows how to generate magnetoacoustic waves 
//at 500 MHz at a fixed field. We define a wave with a spatial 
//and a temporal component as A exp(kx-wt).
//The spatial component is defined with a ScalarMask 
//which can then be added with the temporal component.
// MuMax3: https://mumax.github.io/
// Grid properties
CellSize:=10.7e-9 // 500 MHz wavelength
NumCells:=512
ZNumCells:= 4
SetGridSize(NumCells, NumCells, ZNumCells)
SetCellSize(CellSize, CellSize, 18.5e-9) // 74-nm thick
SETPBC(6,6,0)
Setgeom(universe())

//MATERIAL PARAMETERS FOR Fe3Si
Msat=955e3
Aex=1e-11  
alpha = 0.005
// Cubic Anisotropy
anisC1=vector(1,1,0) 
anisC2=vector(-1,1,0)
Kc1=-4400
Kc2=0
// Magnetoelastic constants
B1=5e6 
B2=14e6 
tmax:=5e-7

freq:=500e6 // SAW frequency
lambd:=1. //in sample width units

// Initial magnetization state in Easy Axes
angle := Pi/2.
m = uniform(cos(angle), sin(angle), 0)

// Applied magnetic field with a small misalignment of 0.3 degrees
hh:=6e-3
angleB := Pi/4. + 0.3*Pi/180. 
B_ext=vector(hh*cos(angleB), hh*sin(angleB), 0)

defregion(10,Cell(0,0,0))
defregion(11,Cell(NumCells/2,NumCells/2,0))

//SAW
angleSAW := Pi/4.
epXX := 7.e-5
epXZ := 1.e-7
epZZ := 3.e-5

// Mask (Spatial component of SAW)
scx := NewScalarMask(NumCells, NumCells, ZNumCells)
ssx := NewScalarMask(NumCells, NumCells, ZNumCells)
scz := NewScalarMask(NumCells, NumCells, ZNumCells)
ssz := NewScalarMask(NumCells, NumCells, ZNumCells)
for i:=0; i<NumCells; i++{
	for j:=0; j<NumCells; j++{
		exc := cos( 2*Pi*(i+j)/(NumCells*lambd))
		exs := sin( 2*Pi*(i+j)/(NumCells*lambd))
		for k:=0; k<ZNumCells; k++{
			scx.setScalar(i,j,k,exc)
			ssx.setScalar(i,j,k,exs)
			scz.setScalar(i,j,k,exc)
			ssz.setScalar(i,j,k,exs)
		}
	}
}
exx.add(scx, epXX * pow(cos(angleSAW),2) * cos(2*Pi*(freq*(t))))  
exx.add(ssx, epXX * pow(cos(angleSAW),2) * sin(2*Pi*(freq*(t))))
eyy.add(scx, epXX * pow(cos(angleSAW),2) * cos(2*Pi*(freq*(t))))
eyy.add(ssx, epXX * pow(cos(angleSAW),2) * sin(2*Pi*(freq*(t))))
exy.add(scx, epXX * cos(angleSAW)*sin(angleSAW) * cos(2*Pi*(freq*(t))))
exy.add(ssx, epXX * cos(angleSAW)*sin(angleSAW) * sin(2*Pi*(freq*(t))))

exz.add(ssx, epXZ  * cos(angleSAW)* cos(2*Pi*(freq*(t))))
exz.add(scx, -epXZ * cos(angleSAW)* sin(2*Pi*(freq*(t))))
eyz.add(ssx, epXZ  * sin(angleSAW)* cos(2*Pi*(freq*(t))))
eyz.add(scx, -epXZ * sin(angleSAW)* sin(2*Pi*(freq*(t))))

ezz.add(scz, -epZZ * cos(2*Pi*(freq*(t))))
ezz.add(ssz, -epZZ * sin(2*Pi*(freq*(t))))

tableadd(m.region(10))
tableadd(m.region(11))
tableautosave(1e-11)
relax()
run(tmax)

\end{Verbatim}